\documentclass[
aps,%
12pt,%
final,%
notitlepage,%
oneside,%
onecolumn,%
nobibnotes,%
nofootinbib,% 
superscriptaddress,%
noshowpacs,%
centertags]%
{article}

\begin{document}

\title{
       %Revision on the magnetic dipole model revision              
       {\Large
       On "Pulsar dynamics: magnetic dipole model revisited"
       }
      }

\author{ 
        { 
	 D. P.\ Barsukov,\ 
	 E. M.\ Kantor,\ 
	 A. I.\ Tsygan
	 \footnote{{\it e-mail}:\ tsygan@astro.ioffe.rssi.ru} $\,$,	
        } \\
        {\it \small Ioffe Physical Technical Institute,}\\  
        {\it \small Politekhnicheskaya 26, 194021 St. Petersburg, Russia }
        \\
        %{\rm Key words. stars: neutron -- dense matter}
       }
\date{${}$} 
\maketitle

On November the 17th an article written by G. Lugones and I. Bombaci 
appeared in astro-ph/0411507. The authors reconsider a pulsar braking 
due to magneto-dipolar deceleration.
The authors made remark that it is necessary to take into account 
the time dependence of angular velocity $\Omega$ while differentiating 
phase $\phi=\Omega t$ by time, so $\dot{\phi}=\dot{\Omega}t+\Omega$.
%(note that their definition of the angular velocity is not conventional, 
%usually the angular velocity is defined as $\Omega=\dot{\phi}$
%or $\phi=\int \Omega dt$).
But their definition of the angular velocity is not conventional, 
usually the angular velocity is defined as $\dot{\phi}$. 
Their analysis shows that the polar magnetic field strength $B_p$ 
inferred from timing observations may be up to a factor 4 larger 
than currently expected and braking index is also different. 
This result is not correct.  
They did not use this expression ($\dot{\phi} =\dot{\Omega}t+\Omega$) 
while calculating the rotational energy of the pulsar 
$E=I\dot{\phi}^{2}/2$ 
(the formula which was used by the authors is $E=I \Omega^{2}/2$).   
\\
Use of correct formula for the rotational energy leads 
to the following differential equation for the pulsar deceleration:
\begin{equation}
\dot{\phi}\ddot{\phi}= -K\left(
                         \dot{\phi}^{4}+\ddot{\phi}^{2}
                         \right)
\nonumber
\end{equation}
Neglecting the second term (which is small in comparison to the first one) 
in the right hand side of this equation, 
we can rewrite it in traditional form:
\begin{equation}
\ddot{\phi}= -K\dot{\phi}^{3}
\nonumber
\end{equation}
and we arrive at the conventional values for the magnetic fields and 
braking index equal to 3.

\end{document}